\documentclass{article}
\usepackage{waspaa21,amsmath,graphicx,url,times}
\usepackage{color}

\usepackage{hyperref}
\usepackage{amsfonts}
\title{Point Cloud Audio Processing}

\name{Krishna Subramani$^{\sharp}$, Paris Smaragdis$^{\sharp,\flat}$}
\address{$^{\sharp}$University of Illinois at Urbana-Champaign, $^{\flat}$Adobe Research}

\begin{document}

\ninept
\maketitle

\begin{sloppy}

\begin{abstract}
Most audio processing pipelines involve transformations that act on fixed-dimensional input representations of audio. For example, when using the Short Time Fourier Transform (STFT) the DFT size specifies a fixed dimension for the input representation. As a consequence, most audio machine learning models are designed to process fixed-size vector inputs which often prohibits the repurposing of learned models on audio with different sampling rates or alternative representations. We note, however, that the intrinsic spectral information in the audio signal is invariant to the choice of the input representation or the sampling rate. Motivated by this, we introduce a novel way of processing audio signals by treating them as a collection of points in feature space, and we use point cloud machine learning models that give us invariance to the choice of representation parameters, such as DFT size or the sampling rate. Additionally, we observe that these methods result in smaller models, and allow us to significantly subsample the input representation with minimal effects to a trained model performance.\footnote{\small Code: https://github.com/SubramaniKrishna/point-cloud-audio}
\end{abstract}


\section{Introduction}
\label{sec:intro}

Conventional machine learning models take as input fixed dimensional inputs. For audio, the most common representations we use are directly derived from the Fourier domain. To get, for example, the STFT, we fix parameters like the DFT size and sampling rate a priori based on the task. For a DFT transform size of $N$, the magnitude spectrum feature vector will belong to $\mathbb{R}^{N/2+1}$. What happens if we then change the transform size to $N^{'} \neq N$? A conventional feed-forward neural network would not be able to directly process this differently-sized vector. Consider the task of building a system where we cannot control the DFT sizes or sampling rates of the data we collect and train on. With current methods, we would be forced to either discard incompatible data or resample it, neither being a desirable solution. Models that can process data with varying representations would allow us to collect from and deploy to a broader set of situations, and would facilitate dynamically adjusting representations to satisfy various constraints (e.g. reducing network bandwidth by adjusting an edge device's sampling rate, or reducing computations and memory usage by subsampling the feature space).


Instead of processing fixed length vectors, we can work with collections of points in feature space. More formally, instead of using a feature vector representation $\mathbf{x} = [x_1, x_2, ..., x_N]$, we can use a set of points $\mathbf{p}_i$ forming a set $\mathbb{P}$ that explicitly define each value in the feature space using lower dimensional vectors. One way to do that would be using a set of two-dimensional vectors, $\mathbf{p_i} = [i,x_i]$, which would independently encode each feature value with its index.  Although this particular case is an inefficient representation, this encoding can be more powerful since it can enable the encoding of non-integer feature spaces (e.g. a Hz value instead of a frequency bin index) and it allows us to subsample at will.  More importantly, it can allow us to escape from the fixed-size feature vector mentality. We call a set of such points a point cloud as is common in the computer vision literature. A defining characteristic of point clouds is that they are permutation invariant (i.e., the ordering of points does not matter). Thus, systems designed to process point clouds must also be permutation invariant, and furthermore, must also be invariant to the number of the points taken into consideration. One proposition to deal with point clouds is to quantize them by fitting them onto a regular grid, and using the same methods as we currently do. The problem with this is that we are unnecessarily increasing the amount of data we need, and also possibly losing resolution information by quantizing onto a regular grid. This forms our motivation for this paper: to design network architectures that exploit the invariant structure of audio point clouds and operate directly on them.

The use of point clouds in machine learning is not new, PointNet and PointNet++ \cite{qi2016point,qi2017pointnet++} are networks for classification and segmentation that directly operate on point cloud data from, e.g., LIDAR sensors. The main idea is to operate on points independently and identically, and then aggregate information across them (to ensure invariance). Through a combination of these operations, the network learns to select points that are relevant in the optimization of the downstream task. Convolutional Networks \cite{kipf2016semi}, Dynamic Edge Convolutional Networks \cite{wang2018dynamic}, PointGCN \cite{zhang2018graph} and PointCNN \cite{li2018pointcnn} use Graph based approaches to directly learn features from point clouds.

Deep Sets \cite{zaheer2017deep} develops a framework for machine learning tasks that operate directly on sets. The authors claim that any permutation invariant function can be decomposed in the form $\rho \left(\sum_{x \in \mathcal{X}} \phi(x) \right)$ where $\rho,\phi$ are point-wise learnable networks. However, Deep Sets does not take into account inter-point interactions since the inner summation neglects interactions. This issue can be addressed by using a Transformer motivated architecture. The Transformer as implemented in Vaswani et al.  \cite{vaswani2017attention} is permutation invariant if you remove the positional encodings. Inspired by this, Lee et al. \cite{lee2019set} develop Set Transformers that directly operate on set structured data.

In our work, we build on top of Set Transformers by using them to classify audio point clouds. We can represent audio spectra as point clouds of (Frequency, Magnitude) or (Time, Frequency, Magnitude) points. We begin by contrasting a Set Transformer network with conventional feed-forward classifiers for audio frame classification. We then present a series of experiments that walk through how our proposed model achieves invariance to the choice of window size and sampling rate. We finally extend the task to take into account the temporal ordering of audio and classify spectrograms instead of spectral frames.




\section{Point Cloud Processing}
Here, we will briefly explain Set Transformers \cite{lee2019set}. The main building block in Set Transformers is the \textit{Multihead Attention Block} (MAB) which computes \textit{attention} between two input point clouds $\mathbb{X}$ and $\mathbb{Y}$. Attention is computed as $\rm{softmax}(\mathbf{Q}\cdot\mathbf{K}^\top)\cdot\mathbf{V}$ where $\mathbf{Q},\mathbf{K},\mathbf{V}$ represent projections of available points via learned transforms. This is directly inspired from the Transformer encoder in Vaswani et al. \cite{vaswani2017attention} and skips the positional encoders and regularizers. 
Following that, a \textit{Set Attention Block} (SAB) uses the MAB to compute \textit{self-attention} for the input set. Thus, for a $d$-dimensional input point cloud $\mathbb{X}$ with $n$ points, $\rm{SAB}(\mathbb{X}) = \rm{MAB}(\mathbb{X},\mathbb{X})$. To take into account higher-order interactions, multiple SAB blocks are often sequentially applied. The output of each SAB block is a point cloud of a user-defined latent dimension.
Another important thing to consider computationally is that self-attention is quadratic in the number of input points (which can be prohibitively expensive for large collections of points). Thus, \textit{Induced Set Attention Blocks} (ISAB) are proposed where $k << n$ inducing points are learned from the data (akin to subsampling the point cloud). These points are learned when optimizing w.r.t the downstream task (classification in our case). Thus, if $\mathbb{I}$ is the smaller collection of points that are ``learned'', the ISAB block is $\rm{ISAB}(\mathbb{X}) = \rm{MAB}(\mathbb{X},\rm{MAB}(\mathbb{I},\mathbb{X}))$. This reduces the computational cost from $\mathcal{O}(n^2)$ to $\mathcal{O}(k\cdot n)$. The final block in the Set Transformer is a pooling block to aggregate information from the intermediate representations (to ensure permutation invariance). Set Transformers do this with pooling by a \textit{Multihead Attention Block} (PMA) which uses self-attention to aggregate information from all the points into a final $k \times D$ representation. For our work, we use $k = 1$, thus effectively reducing the point cloud to a $D$-dimensional representation.

\subsection{Frame-wise Classification}
Let us now consider how we can use these ideas to make a sound recognition system that does not rely on a fixed sampling rate, or a fixed-size feature vector.
\subsubsection{Toy Example}
\label{sec:framewise}
We first consider the task of classifying single spectral frames of different audio classes. As a baseline, we will use a standard two-layer feed-forward network classifier with a hidden layer size of 8 units and an input size of 64.  Whereas for the point cloud representation we will use a Set Transformer with 2 ISAB and one PMA with hidden dimension 2, followed by linear classification layer outputting 2 class probabilities.
\begin{figure}
    \centering
    \includegraphics[scale = 1.6]{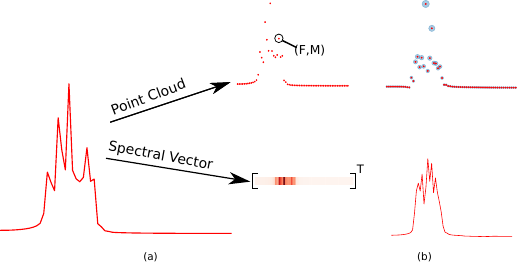}
    \caption{(a) Point cloud vs. vector representation. The underlying continuous spectrum is represented as a fixed-size spectral vector with a quantized frequency axis (bottom), vs. a collection of frequency/magnitude points in a real-valued frequency space (top).
    (b) Learned representations for class 1. The hidden layer of the feed-forward model learns to correlate strongly with that class' spectra (bottom), whereas the learned attention model learns to focus on frequencies in that class' distinguishing frequency band (top).}
    \label{fig:toyexample_input}
\end{figure}
To illustrate the difference in how a feed-forward network and Set Transformer process the input, we construct a simple toy example to classify two simple signals. The signals are band-pass filtered white noise such that the frequency bands for the two classes are exclusive.
\autoref{fig:toyexample_input} (a) shows the input to the two respective models. For the Set Transformer, we will represent the input as a point cloud of two-dimensional vectors $\mathbf{x}_i=[f_i,m_i]$ that contain a frequency $f_i$ (in Hz) and its corresponding magnitude $m_i$ from the input spectrum, whereas the baseline model will process the magnitude DFT vector directly.

\autoref{fig:toyexample_input} (b) shows the learned attention values for one class from our Set Transformer. The radius of the blue circle is proportional to the attention that point gets. We see that after training, the Set Transformer learns to attend to points inside a specific frequency band, and passes that information upstream. For the linear feed-forward model we show one of the learned weight matrix columns after training. As expected, it learns a template for the frequency structure of one class, and it propagates information ahead by computing an inner product between the input vector and multiple such templates, which it aggregates and uses for classification in the final layer.

Since in the baseline we compute an inner product with fixed dimensional templates, we do not have the flexibility to change the size of the input vector. Instead, the Set Transformer learns to ``attend'' to the important frequencies regardless of how many points we use and how we sampled them.  Thus, we can see the important advantages of this approach.  Because frequency is encoded as a real-valued variable with each point, we can sample the frequency space arbitrarily (i.e., we can use any size DFT or sampling rate that we like since our model does not require quantized frequency values on a predefined bin index grid), and we can subsample the input spectrum selecting only a subset of the available input points thus reducing the amount of data that we need to process.

\subsubsection{ESC-10 Classification}
To examine the efficacy of our model on real-world data, we consider the ESC-10 dataset \cite{piczak2015dataset}, which is a 10-class sound classification set. We preprocess the audio data to remove any boundary silence and split the data into an 80-20 train-test set. We use a training DFT size of $N = 2048$pt, and train a frame-wise baseline (FB) and a frame-wise Set Transformer (FST) model classifier. The FB model is a feed-forward network with two hidden layers of size 512 and 256 and leaky ReLU activations. The FST is a combination of two ISAB and one PMA block with a hidden dimension of 64, followed by a linear layer that outputs 10 class probabilities. Both models are trained to minimize the cross-entropy loss with an Adam optimizer \cite{kingma2014adam} of step size $10^{-3}$ for 500 epochs. To prevent overfitting, we add an $\ell_2$ weight regularizer to the optimizer with  $\lambda = 10^{-3}$. We perform the following experiments to demonstrate the advantages of FST over FB: 1) Varying the input window size and sampling rate, and 2) Subsampling the input spectra.

\begin{figure}[t]
    \centering
    \includegraphics[scale = 0.5]{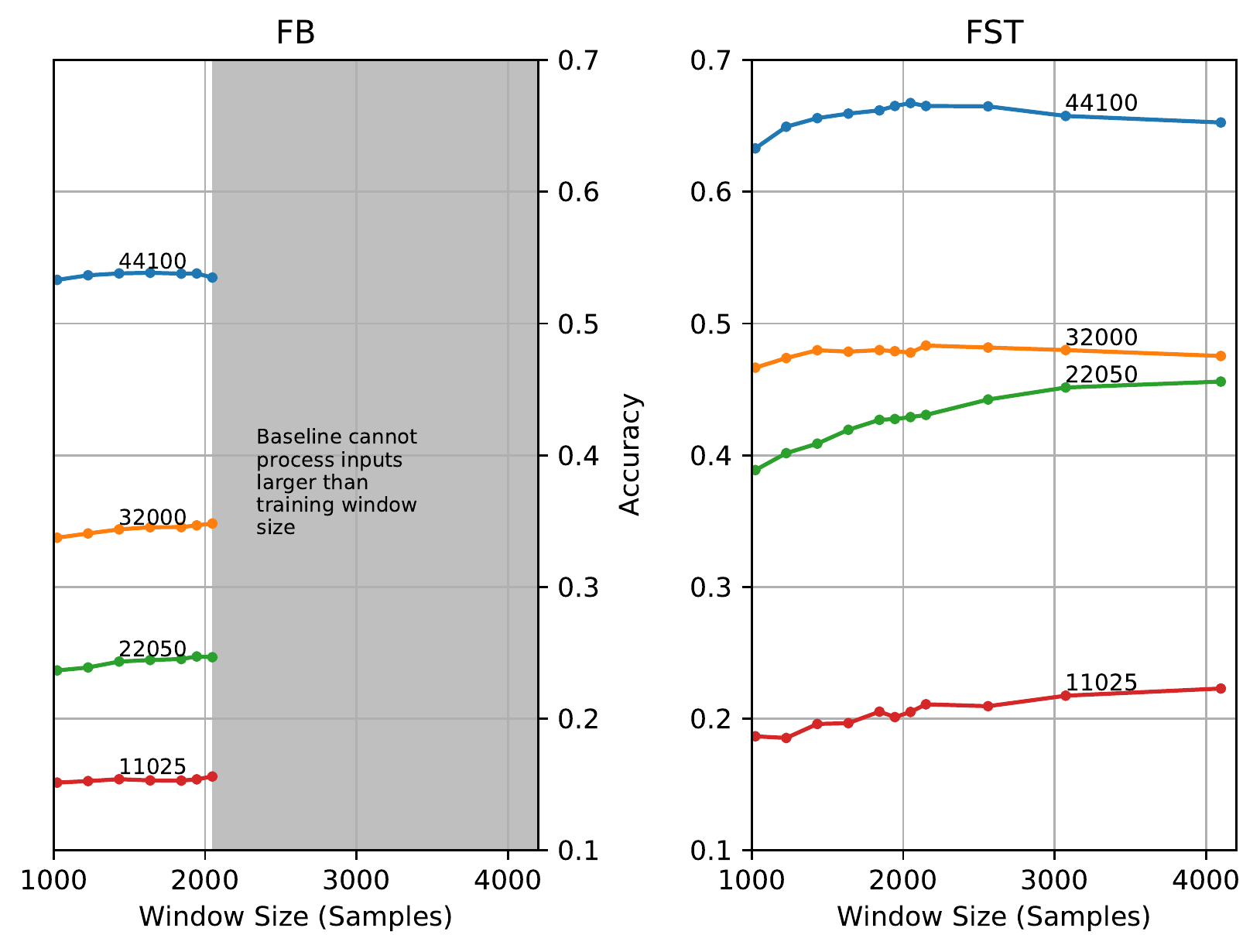}
    \caption{Accuracy vs. varying window size and sampling rate.  On the left we show the results of a baseline modes with a fixed-input representation, and on the right our proposed method.  Note how our method tolerates well changes in the input size or the input sample rate, except for extreme cases.}
    \label{fig:framewise_nfs}
\end{figure}
If we treat our input data as a collection of frequency and magnitude pairs, then changing the DFT size or sampling rate would not cause drastic changes in these values themselves. Thus, we expect the FST to be robust to such transformations. Since the FB model expects an input vector of fixed length, for smaller windows we can zero pad the DFT's input to $N = 2048$ to accommodate the model. However, we will not be able to feed into that model inputs of a larger window size. \autoref{fig:framewise_nfs} shows the variation in classification accuracy with varying window sizes and sampling rates. The first thing to notice is that the FST model performs consistently better than the FB model for each sampling rate. Moreover, the drop in accuracy by changing the sampling rate is more significant for the FB model, indicating that the baseline representation is not as robust to changes in the sampling rate. We also see that the FST model is robust to both increases and decreases in the window length, as compared to the FB model where we can only reduce and not increase the window length. The results from this plot strengthen our belief that treating the input as a point cloud allows the model to learn a representation that can tolerate window size and sampling rate differences from the training data.

Since we now give a set as an input to our model, another question that can be asked is if we can subsample the set and see how that impacts classification (a real-world case would be classification with missing data, or forced subsampling due to computational constraints on edge devices). We try two methods of subsampling: 1) We keep the $K$ points with highest spectral magnitudes and 2) We keep $K$ randomly chosen points. 
We cannot use the same process of dropping points for the FB model because the input has to be fixed length. Thus, we modify the procedure by zeroing out the non-chosen frequency bins (to indicate the lack or absence of that information). To make the FB model robust to this zeroing out, we train using dropout \cite{hinton2012improving} at the input layer of the FB model. \autoref{fig:framewise_subsampling} shows the classification accuracy as a fraction of the input points kept. The random subset experiments are repeated 10 times, and we plot the error-bars for the random sampling experiments. We see that for both the FST and FB models, random sampling performs better than the corresponding top-$K$ sampling. Moreover, we observe that the error-bar magnitudes amongst different trials (indicating the standard deviation in obtained accuracy) are extremely small, suggesting that random sampling can be a good strategy to down-sample the spectrum with predictable drops in accuracy. We also see a really interesting result with the FST model, we can achieve little drop in accuracy just by keeping $\approx 30\%$ of the points, which is a significant reduction in the data we need to classify the input. From the same plot, we also see that, for the FST model, as we select more of the largest $K$ points (equivalent to adding more coordinates from the spectral vector), the accuracy increases gradually, but for the FB, we observe a sudden jump from very low to very high accuracy when we keep $95\%$ of the points. This strengthens our belief that the Set Transformer model actually learns to process feature space coordinates independently, and as you add more information (in terms of more coordinates), the accuracy increases. This is in contrast to the FB case, which treats the input as a vector, and thus expects most of the information to be present before the model can make any decision (indicated by the sudden flip from poor classification to good classification when most of the points are present).
\begin{figure}
    \centering
    \includegraphics[scale = 0.5]{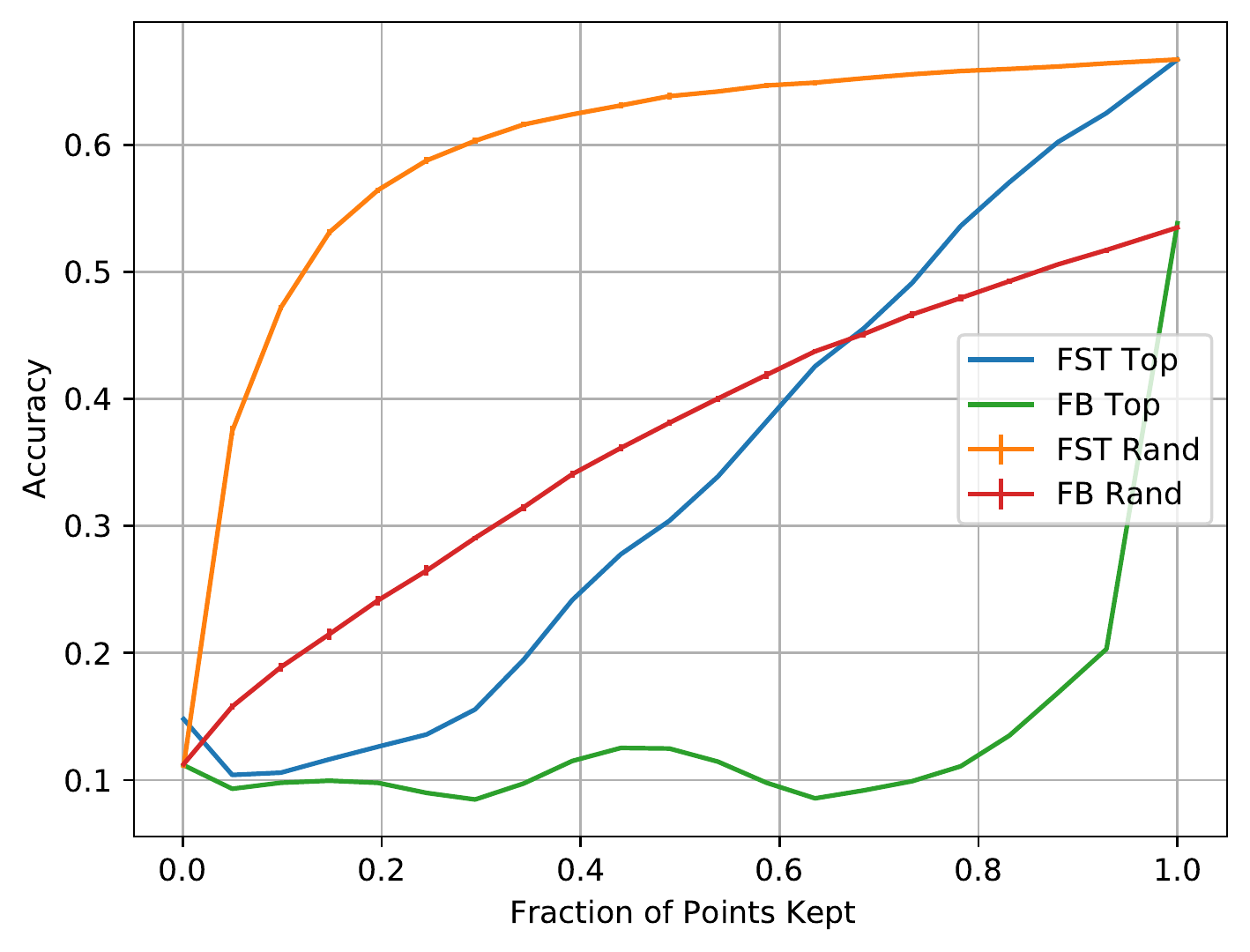}
    \caption{Accuracy vs. Fraction of input features kept.  The lines labelled ``Top'' keep the largest magnitudes available, whereas the lines labelled ``Rand'' randomly subsample the input space.  Note how the FST model is more robust to input subsampling, and how random sampling seems to not influence its performance significantly until a very low value.}
    \label{fig:framewise_subsampling}
\end{figure}
\subsection{Spectro-Temporal Processing}
We now extend to the task of classifying time-frequency data. The equivalent problem in Point Cloud literature is the task of learning from spatio-temporal Point Clouds. PointLSTM \cite{min2020efficient} proposes an LSTM architecture that acts directly on Point Clouds by combining past information with the present from a temporal Point Cloud sequence. Cloud LSTM \cite{zhang2019cloudlstm} introduces a Dynamic Point Cloud Convolution to aggregate information across time from the Point Cloud. Since these models did not perform better, instead of utilizing an LSTM for modeling time we expanded the point cloud representation to include time explicitly. We appended a time coordinate to our 2D point cloud and made it a 3D point cloud that uses points $\mathbf{x}_i = [t_i,f_i,m_i]$ that are tuples of time $t$, frequency $f$ and magnitude $m$. We call this the 3D Set Transformer (3ST). For an equivalent baseline, we compare with a CNN using the same temporal receptive field as the Set Transformer model. The CNN model uses 2D convolutions with kernel size (10,1) and unit stride, followed by a simple classification layer. The 3ST is a combination of 2 ISAB and one PMA block, all of hidden dimension 64, followed by a classification layer. Both models are trained to minimize the cross-entropy loss with an Adam optimizer using a step size of $10^{-3}$ for 500 epochs, and we use the same $\ell_2$ regularization weight as the previous experiments. The inputs to both our models are 10-frame spectrograms using a window size of $N = 1024$pt.
\begin{figure}[t]
    \centering
    \includegraphics[scale = 0.5]{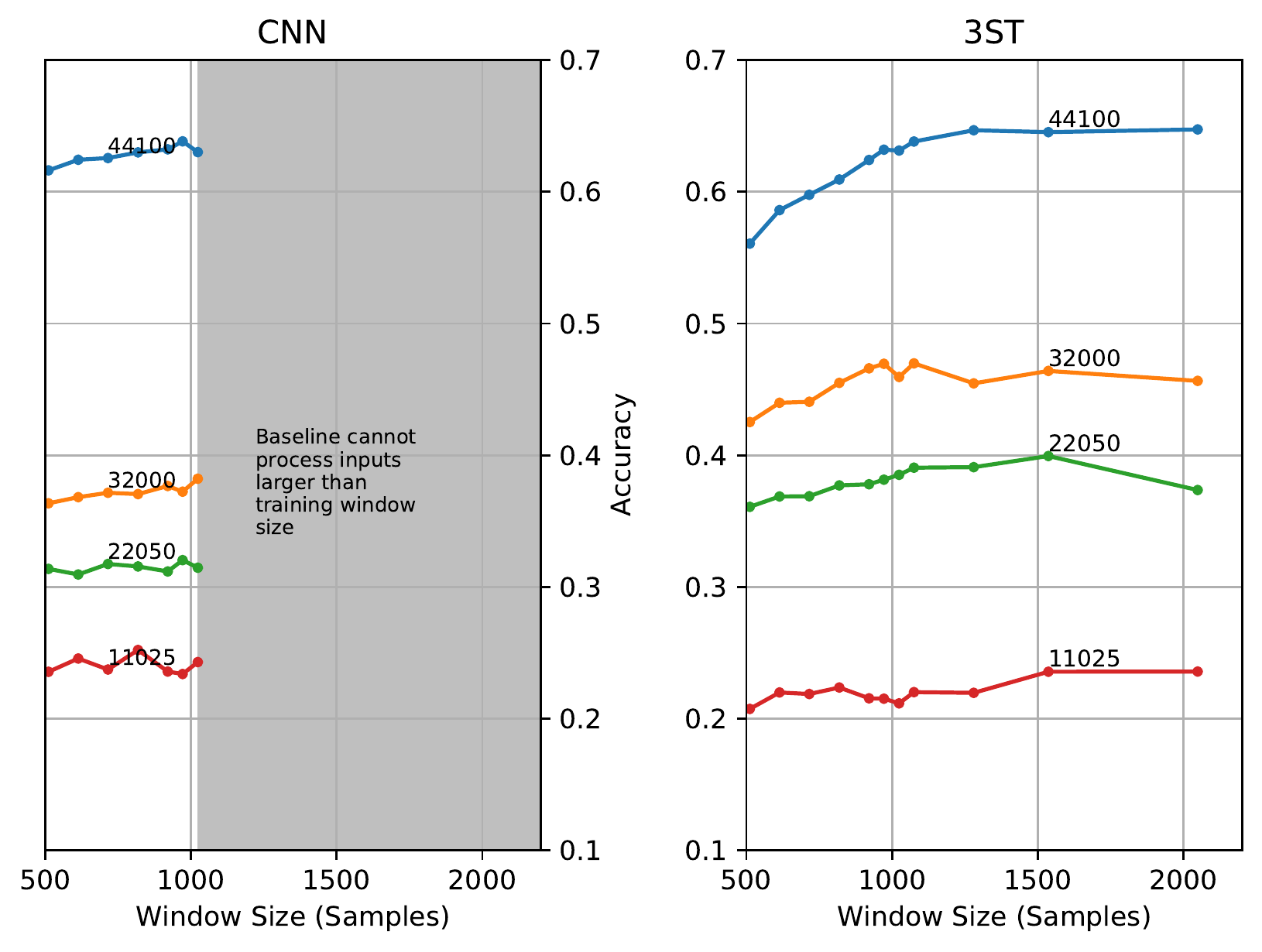}
    \caption{Accuracy vs. varying window size and sampling rate. As before, the proposed representation shows resilience to variations from the training representation.}
    \label{fig:temporal_nfs}
\end{figure}

We perform the same experiments as we did for the frame-wise case. \autoref{fig:temporal_nfs} shows accuracy variation with varying window size and sampling rate. As previously, the 3ST model performs gracefully when presented with varying representations, although it seems to have more difficulties with smaller windows. \autoref{fig:temporal_subsampling} shows accuracy variation with the fraction of features kept. As before, random sampling presents a promising strategy to down-sample the point cloud with marginal drops in accuracy. We can randomly subsample to $\approx 10 \%$ of the size, and see a very small drop in accuracy compared to using the entire point cloud.

We hypothesize that sampling top magnitude points only (indicating presence of energy) does not provide information to discriminate between classes. We would also have to know where energy is absent (otherwise we would not be able to tell wideband noise from a rich harmonic sound). Randomly sampling uniformly from the entire spectra would have a high likelihood of some sampled points lying in areas that might help in classification. This is corroborated by the fact than when we selectively sample from high gradient areas in the spectrotemporal space we get results which are very close to the random sampling case, implying that knowing where there is no energy is important.

\begin{figure}[ht]
    \centering
    \includegraphics[scale = 0.5]{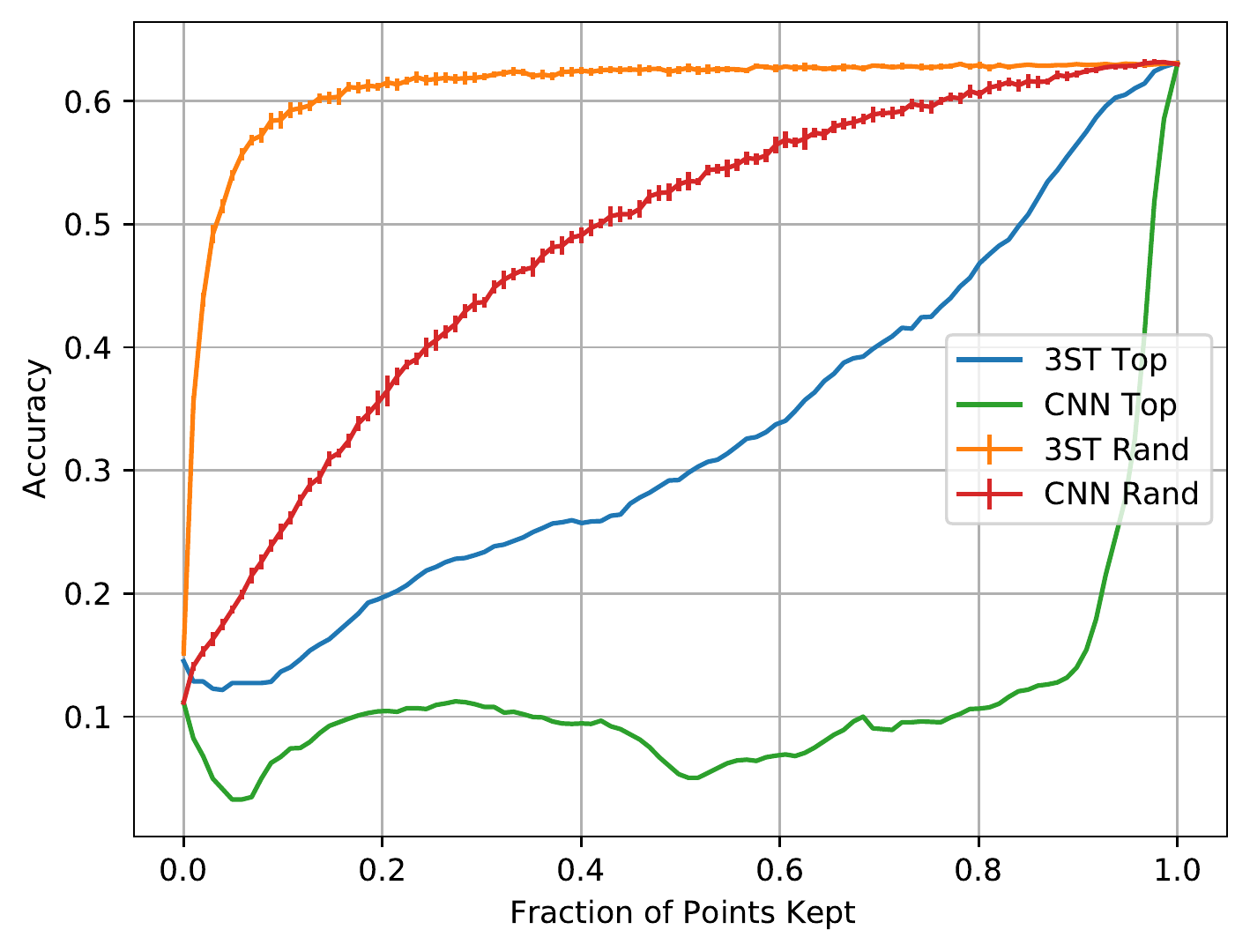}
    \caption{Accuracy vs Fraction of points kept. Note how the proposed 3ST model is highly robust to random subsampling of the input representation, and significantly more so than the baseline.}
    \label{fig:temporal_subsampling}
\end{figure}

\subsection{Network Size and Operation Count}
To enforce competitiveness in our experiments, we ensure that the number of trainable parameters in our Set Transformer models (FST and 3ST) are fewer than the number of trainable parameters for the respective baseline models. Our FST model contains 80,202 trainable parameters, and the FB model contains 660,492 parameters. Our 3ST model contains 80,394 trainable parameters, and the CNN contains 158,049 trainable parameters. Thus, in both cases, the point cloud models show better (or similar) performance with a significant reduction in the number of trainable parameters. However, this is achieved at an increase in the number of operations per iteration. The FST model requires approximately two orders of magnitude more operations per iteration than the baseline model, and the 3ST requires three orders of magnitude more operations per iteration. This manifests itself in longer training times for the Set Transformer models. From our observations in the subsampling experiments, we see that we can dramatically subsample the input and achieve similar performance. We can also do this during training i.e. we can subsample the point cloud to a 16th of the original size, and we observe marginal reduction in the accuracy (1\% drop). We do not have to increase the number of epochs for the model to converge. In the same 500 epochs, both the FST and 3ST models converge with the subsampled inputs (and the Set Transformer models take half the time to train compared to the using the entire point cloud). Our training subsampling strategy also leads to an order of magnitude reduction in the number of operations per iteration for both models. In the edge-computing setup, this actually presents a really promising strategy -- we do not need information from the entire input spectrum, we can train with a subset of spectral bins (randomly) chosen, and achieve similar performance as if training with the entire input.

\section{Conclusions}
We introduce a novel scheme for processing audio in the spectro-temporal domain by treating it as a set of points instead of a fixed-dimension array. Through experiments, we demonstrate relative invariance to the choice of processing parameters like the DFT window size and the sampling rate. We also present the added advantages of smaller models in terms of trainable parameters, and the ability to significantly subsample the input representation with marginal performance drop. Finally, we introduce a subsampling strategy during training to reduce the training time. Such a model allows us to infer and learn from varying audio sources without needing to enforce uniform sampling and analysis settings, and we see this as being a big advantage when it comes to real-world deployment.

\bibliographystyle{IEEEtran}
\bibliography{refs21}
%
%
%
%
%
%
%
%
%

\end{sloppy}
\end{document}